\documentclass[aps, prl, twocolumn, showpacs,
amsmath, amssymb, notitlepage, superscriptaddress, floatfix]{revtex4}
\usepackage{epsfig}
\usepackage{graphicx}
\usepackage{psfrag}

\newcommand{\refdot}[1]{Ref.~\onlinecite{#1}}
\newcommand{\refsdot}[1]{Refs.~\cite{#1}}
\newcommand{\figdot}[1]{Fig.~\ref{#1}}
\newcommand{\parref}[1]{(\ref{#1})}

\newcommand{\vect}[1]{\mathbf{#1}}
\newcommand{\units}[1]{\,\ensuremath{\mathrm{#1}}}
\newcommand{\decexp}[1]{\cdot 10^{#1}}
\newcommand{\rrprime}{|\vect{r} - \vect{r}'|}
\newcommand{\angs}{\textrm{\AA}}
\newcommand{\mypsi}{f}
\newcommand{\mysection}[1]{\textbf{#1}.}
\newcommand{\EE}{\mathcal{E}}

\newcommand{\torr}{Torr}

\newcommand{\cwi}{
CWI, P.O. Box 94079, 1090 GB Amsterdam, The Netherlands
}

\newcommand{\eindhoven}{
Dept. Physics, Eindhoven Univ. Techn., The Netherlands
}

\newlength{\figwidth}
\newlength{\evolheight}
\begin{document}

\title{Photoionization in negative streamers: fast computations and 
two propagation modes}
\author{Alejandro Luque}
\affiliation{\cwi}
\author{Ute Ebert}
\affiliation{\cwi}
\affiliation{\eindhoven}
\author{Carolynne Montijn}
\author{Willem Hundsdorfer}
\affiliation{\cwi}

\date{\today}

\begin{abstract}
Streamer discharges play a central role in electric breakdown of matter
in pulsed electric fields, both in nature and in technology.
Reliable and fast computations of the minimal model for negative
streamers in simple gases like nitrogen have recently been developed.
However, photoionization was not included; it is important in air 
and poses a major numerical challenge.
We here introduce a fast and reliable method to include photoionization
into our numerical scheme with adaptice grids, and we discuss its importance 
for negative streamers. In particular, we identify different propagation 
regimes where photoionization does or does not play a role.
\end{abstract}
\maketitle

Streamers are a generic initial stage of sparks, lightning and 
various other technical or natural discharges \cite{rai1991}.  
More precisely, 
when a high voltage pulse is applied to a gap of insulating matter,
conducting streamer channels grow through the gap. 
Streamer propagation is characterized by a strong field enhancement
at the channel tip. This field enhancement is created by a thin
curved space charge layer around the streamer tip as many computations
show. Such computations are quite challenging due to the multiple
inherent scales of the process. 

Recent streamer research largely concentrates on positive streamers in air 
or other complex gases for industrial applications \cite{vel2000}.
This is because positive streamers emerge from needle 
or wire electrodes at lower voltages than negative ones \cite{rai1991}.
Natural discharges such as sprites \cite{ger2000}, on the other hand,
occur in both polarities \cite{Williams06}, 
in particular, when they are not attached 
to an electrode and therefore double ended. Photoionization 
(or alternatively background ionization) is essential for positive streamers:
as their tips propagate several 
orders of magnitude faster than positive ions drift in the local field,
a nonlocal photon-mediated ionization reaction is thought to cause
the fast propagation of the positive ionization front.
Negative streamers, on the other hand, have velocities comparable
to the drift velocity of electrons in the local field, therefore
a local impact ionization reaction can be sufficient to explain their 
propagation.  This is why
photoionization in negative streamers has received much less
attention, most recent work concentrating on
sprite conditions with relatively low electric fields \cite{liu2004+06}.


The nonlocal photoionization reaction depends strongly 
on gas composition and pressure \cite{zhe1982}, in particular, it is much more 
efficient in air than in pure gases. Furthermore, in air its
relative importance saturates for pressures 
well below 60 Torr ($\approx$ 0.1 bar), while it is suppressed like
$\approx 60\;{\rm Torr}/p$ at atmospheric pressure and above.
In this paper we study the effects of photoionization 
on the propagation of negative streamers by means of efficient
computations with adaptive grids.

\mysection{Streamer model}
\label{sect:model}
Streamer models always contain electron drift and diffusion, 
space charge effects and 
the generation of electron ion pairs by essentially local impact ionization. 
We will use a fluid model in local
field approximation as described, e.g., in \refsdot{ebe2006,mon2006}.
A numerical code with adaptive grid refinement 
was introduced in \cite{mon2006} to investigate negative streamers 
in pure nitrogen, where photoionization plays a negligible role.  
With this code even streamer branching could be determined accurately
\cite{mon2006}.
On the other hand, in gases like air where photoionization cannot 
be neglected, 
photons emitted from excited molecules can act as a non-local source 
of electron-ion pairs; this has to be included in the computations.
The challenge lies in maintaining computational speed and accuracy
while introducing the nonlocal interaction.  

More precisely, the number of photoionization events 
at a given point $\vect{r}$ results
from integrating the emission $I(\vect{r}')$ of photons at every point
$\vect{r}'$ of the gas volume multiplied by a kernel that contains
an absorption function and a geometrical factor.
The production of photons in air is, on the other hand, proportional to the
number of impacts of free electrons on nitrogen molecules and
hence can be related to the impact ionization $S_i(\vect{r})$.  Thus,
\begin{equation}
  \label{Sph_int}
  S_{ph}(\vect{r}) = 
     \int d^3 \vect{r}' \;\frac{I(\vect{r}')\; f(|\vect{r} - \vect{r}'|)}
	  {4 \pi |\vect{r} - \vect{r}'|^2},\: I(\vect{r}) =  \frac{p_q\xi S_i(\vect{r})}{p + p_q},
\end{equation}
where $\xi$ is a proportionality factor that weakly depends on the
local reduced electric field although it is commonly assumed to be constant
and about $\xi=0.02$.  We must note here that, since the only data
accessible from macroscopic observations is the product $\xi f(r)$,
this is often packed into a single function and called, by a slight abuse of
terminology, absorption function.  In this letter, however, we prefer
to apply this term only to $f(r)$.
The factor $p_q / (p + p_q)$ accounts for the
probability of quenching, i.e.\ for the non-radiative deexcitation of 
a nitrogen molecule due to the collision with another molecule.
The pressure $p_q$ is called quenching pressure and
will be taken here as $p_q = 60$ Torr \cite{Legler62}.  There is
some uncertainty over this value and some authors \cite{liu2004+06, Naidis06,
kul2000-2} prefer $p_q = 30$ Torr.  However, different values of $p_q$ 
within this range affect our quantitative results only marginally 
and our numerical approach and qualitative observations remain unchanged.

Evaluating the integral \parref{Sph_int} numerically in each time step
is very time consuming, since for
each grid point $\vect{r}$ one has to add the contributions of
all emitting grid points $\vect{r}'$. Kulikovsky \cite{kul2000-2}
has assumed cylindrical symmetry and has considered only a relatively 
small number of uniformly
emitting rings, interpolating at finer levels.  
This approximation ignores the small-scale details of the density 
and electric field distributions that matter, e.g., in a branching event.  

\mysection{Numerical implementation of photoionization}
We here present a different numerical method that allows us to keep 
calculating with a locally appropriately refined numerical grid,
and nevertheless to obtain reliable results within decent computing times.
Our approach relies on approximating the absorption
function as 
\begin{equation}
  \label{helm_approx}
  \mypsi(|\vect{r} - \vect{r}'|) = 
    \frac{|\vect{r} - \vect{r}'|}{\xi} \sum_{j=1}^N A_j e^{-\lambda_j |\vect{r} - \vect{r}'|},
\end{equation}
where $\lambda_1\dots \lambda_N$ and $A_1 \dots A_N$ fit 
the experimental data as closely as possible.
This form has the advantage that the integral \parref{Sph_int} 
can be expressed by a set of Helmholtz differential equations for
the $S_{ph, j}$ as
\begin{equation}
  S_{ph} = \frac{p_q}{p + p_q} \sum_{j=1}^N A_j S_{ph, j},~~~
(\nabla^2 - \lambda_j^2)\; S_{ph, j} = S_i,
\end{equation}
with the boundary condition $S_{ph, j}(\vect{r}) \to 0$ far away from
the high field areas.  Thus one now can use the very fast 
algorithms available for
solving elliptic partial differential equations with separable
variables, such as described in \refdot{swe1977} and implemented in
the freely downloadable library FISHPACK. The same algorithm was 
used in \refdot{mon2006} to solve the electrostatic problem \cite{wac2005}.

\begin{figure}
\includegraphics[width=0.9\figwidth]{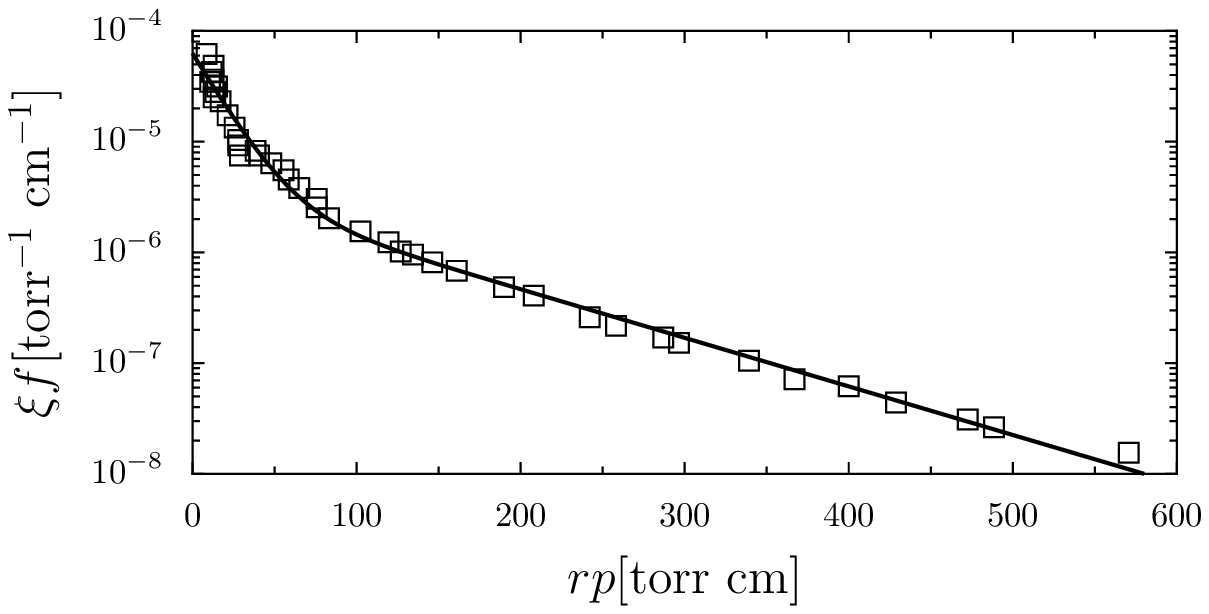}
\caption{\label{air}
  The function $\xi f$ of the photoionizing radiation in
  the range $980-1025 \,\angs$ in air, taken from \refdot{pen1970}
  (squares) and fit according to \parref{helm_approx} (solid line)
  with the parameters 
  $A_1 = 6.0 \decexp{-5} \units{cm^{-1}Torr^{-1}}$, 
  $A_2 = 3.55\decexp{-6} \units{cm^{-1}Torr^{-1}}$, 
  $\lambda_1 = 0.059 \units{cm^{-1}Torr^{-1}}$,
  $\lambda_2 = 0.010 \units{cm^{-1}Torr^{-1}}$.  
}
\end{figure}

For nitrogen-oxygen mixtures like air, the most reliable model for $f$ 
is provided 
by \refdot{zhe1982} based on the experimental measures of \refdot{pen1970},  
despite some recent controversy over these data \cite{pan2005, Naidis06}.
\figdot{air} shows the data for $f$ from \refdot{pen1970}
together with our fit of form (\ref{helm_approx}) with $N=2$.

Note that the asymptotic behavior of \parref{helm_approx} for $\rrprime \to
0$ and $\rrprime \to \infty$ disagrees with that predicted by
\refdot{zhe1982}. Nevertheless, these differences cannot be seen 
in \figdot{air}. For very small distances between the
emitting excited state and the ionized molecule, the
impact ionization is dominant anyway. At distances much larger 
than the largest absorption length $1/(\lambda_jp)$, where most radiation
is absorbed, the identical exponential decay in $r$ dominates 
over the different powers of $r$.

\mysection{Similarity laws}
\label{sect:results}
Without photoionization, there are similarity laws between streamers
at different pressures: They are equal after rescaling lengths, 
times and fields with appropriate powers of the pressure \cite{ebe2006} ---
this generalizes Townsend's historical finding 
that the ratio of electric field
over pressure $E/p$ is the physically determining quantity in a discharge, 
not $E$ and $p$ separately. Photoionization introduces 
a nontrivial pressure dependence through the factor $p_q/(p+p_q)$ 
in (\ref{Sph_int}) and thus breaks the similarity laws 
between streamers at ground level and those in the high altitude, low
pressure regions where sprites appear \cite{liu2004+06}.

\mysection{simulation setup} We have incorporated photoionization into
the numerical code of
\refdot{mon2006} as described above. Air was approximated as an
oxygene-nitrogen mixture in the ratio 20:80. In order to study 
the effect of photoionization on streamer propagation at different
background electric fields, we used fields of 
$100 ~(p / p_0) \units{kV/cm}$ and of $40~ (p / p_0) \units{kV/cm}$
where $p_0$ is atmospheric pressure.
Furthermore, we studied three pressure regimes, namely 
atmospheric pressure ($760\units{\torr}$) and $0.05\units{\torr}$, 
which corresponds to the pressure of the atmosphere at
around $70 \units{km}$ above sea level, where sprites are commonly observed, 
and also the case without any photoionization, 
which corresponds to the physical
limit of very high pressures, when all excited states are rapidly quenched,
or to the case of pure nitrogen.

The length of the computational domain was $4.7\units{mm}/(p/p_0)$
for the higher and $9.4 \units{mm}/(p/p_0)$ for the lower
electric field. The radial extension was large enough that 
the lateral boundaries did not influence phenomena.
At the cathode we imposed homogeneous Neumann boundary
conditions, roughly equivalent to a free electron inflow into the
system.  An initial ionization seed was introduced near the cathode 
as an identical Gaussian density distribution for electrons and ions 
with a maximum of $8.2 \cdot 10^3 / (p / p_0)^3 
\units{mm^{-3}}$ and a radius of $23 (p_0 / p) \units{\mu m}$.

\begin{figure}
\begin{minipage}[b]{0.5\evolheight}
 \includegraphics[width=0.59\evolheight]{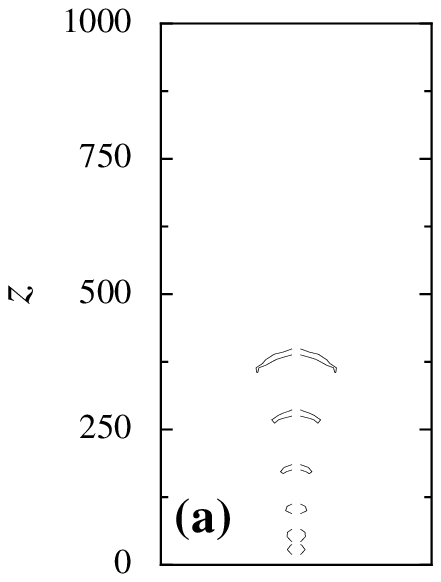} \\
 \includegraphics[width=0.59\evolheight]{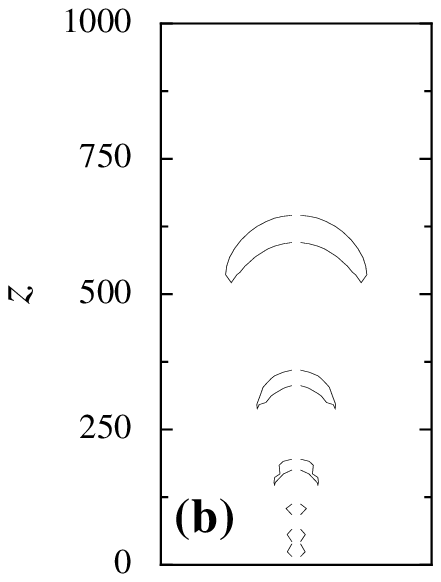}\\
 \includegraphics[width=0.59\evolheight]{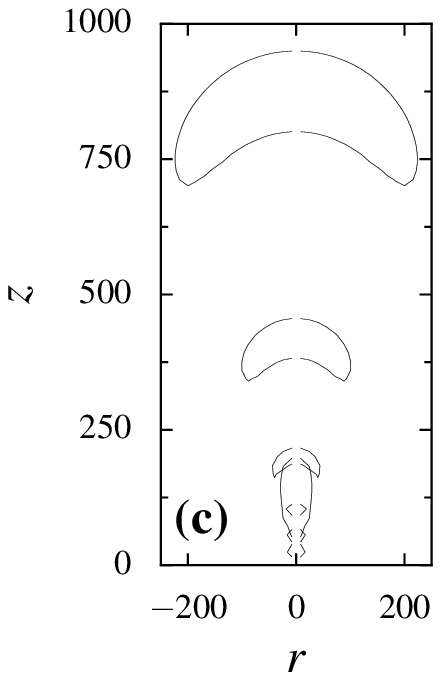}
\end{minipage}
\hspace{0.7cm}
 \includegraphics[height=2.5\evolheight]{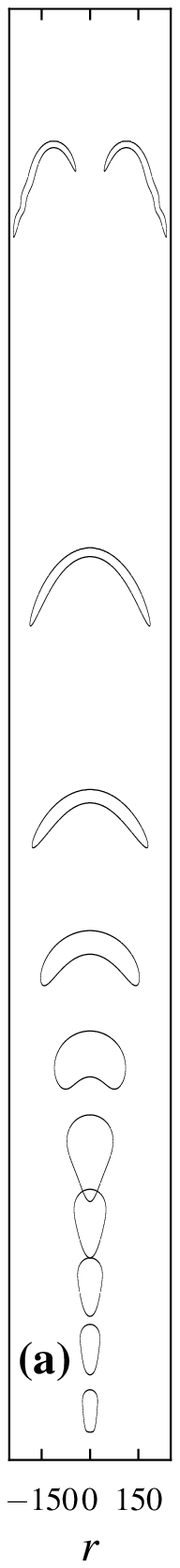} 
 \includegraphics[height=2.5\evolheight]{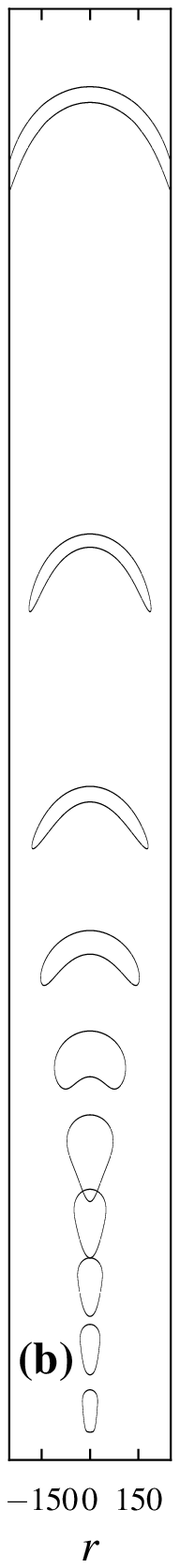}
 \includegraphics[height=2.5\evolheight]{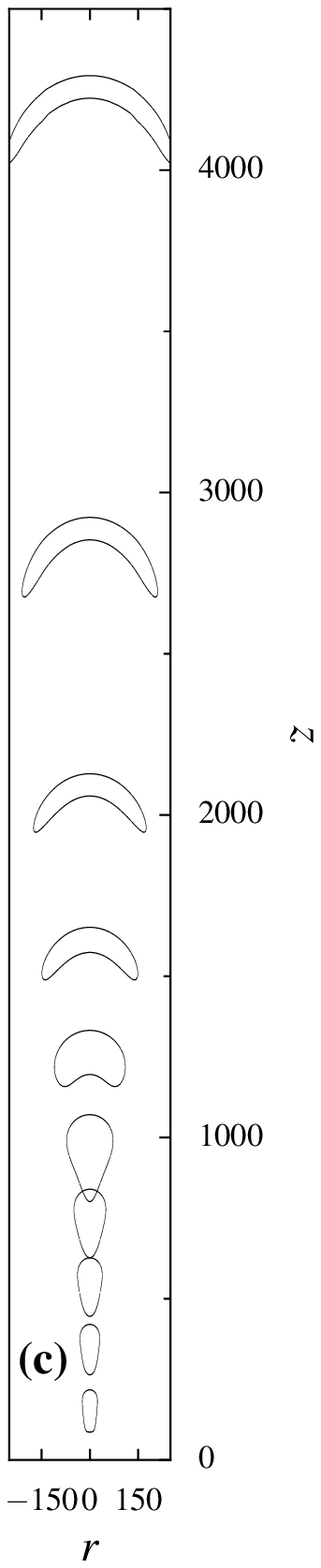}
 \caption{\label{low-e}
   Evolution of streamers in a field of $\EE=100$ kV/cm $p/p_0$ 
   (left column) and in a field of $\EE=40$ kV/cm $p/p_0$
   (other three columns). Plotted is
   the contour of the half-maximum of the space charge at different
   times. The time interval between two consecutive snapshots is
   $\Delta t = 150$ ps / $(p/p_0)$ for the high field and
   $\Delta t = 2400$ ps / $(p/p_0)$ for the low field. Lengths 
   are measured in units of $2.3$ $\mu$m  / $(p/p_0)$.
   Shown are streamers (a) without photoionization, 
   (b) in air at atmospheric pressure, 
   and (c) at low pressures ($p\ll 60 \units{\torr}$). Note that the 
   computational domain is larger than the plotted area.}
\end{figure}

\mysection{Simulation results and conclusions}
Some simulation results for the evolution of the streamer head 
in different fields and pressures are shown in
Fig.~\ref{low-e}. Let us focus first on the high field 
regime which is represented in the left column of the figure;
there it can be seen that during the first three to four time steps,
the streamer development is barely affected by photoionization processes.  
However, eventually a new phase sets in where the streamer 
accelerates significantly. This acceleration is the stronger,
the higher the relative contribution of photoionization, i.e., 
the lower the pressure.  On the other hand, field enhancement is much
weaker: it increases by $\sim 400\%$ without photoionization and only
by $\sim 60\%$ in the low pressure case.

In the lower field case, a very different behavior is seen: 
photoionization hardly
changes the streamer velocity. However, now it does suppress
streamer branching as also found in \refdot{liu2004+06}.
This can directly be related to the fact that photoionization
makes particle distributions smoother, and that a smoother
space charge layer is less susceptable to a Laplacian instability 
\cite{arr2002,ebe2006}.

\begin{figure}
\includegraphics[width=\figwidth]{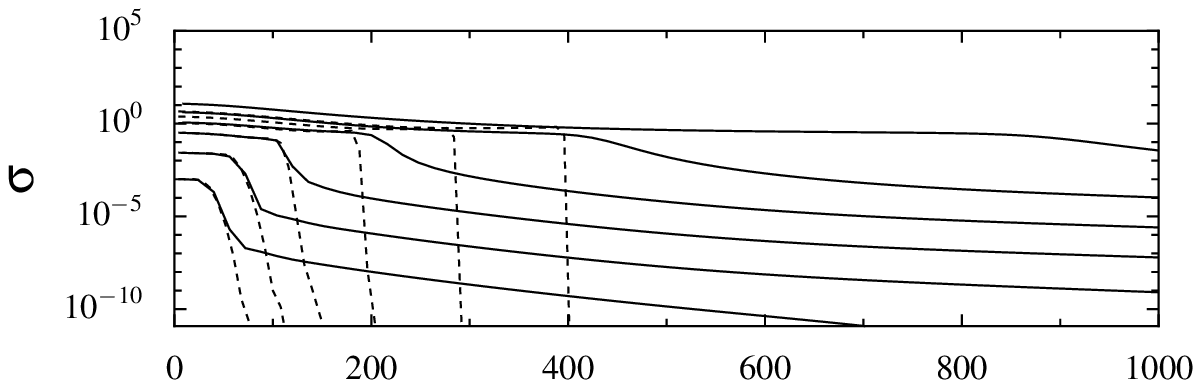}
\includegraphics[width=\figwidth]{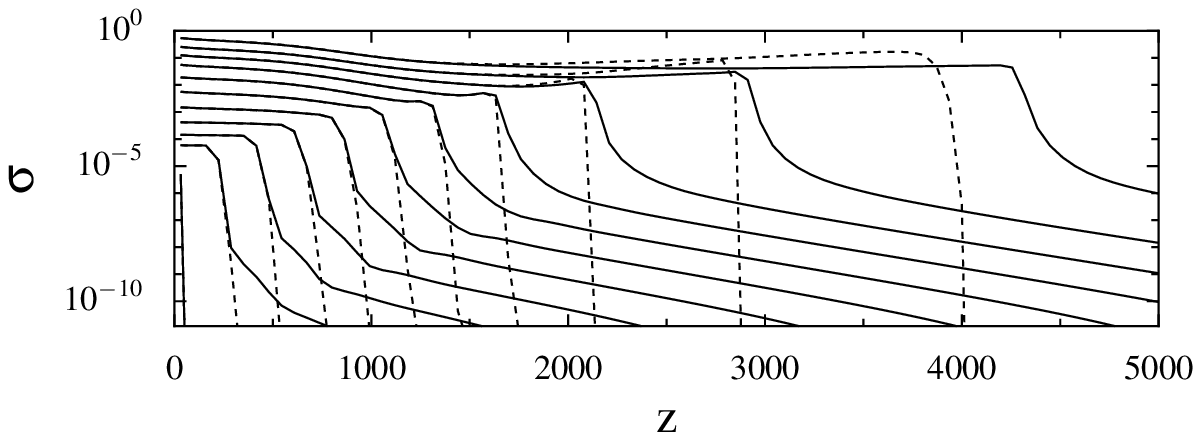}
\caption{\label{airXX_neg_cmp-sigma}
  Evolution of the logarithmic electron density on the streamer axis
  at pressure $p \ll 60 \units{\torr}$ (solid lines) and without
  photoionization (dashed lines) in the high field (above), corresponding
  to columns 2 and 4 in \figdot{low-e},  and low field (below), which
  corresponds to the upper and lower plots in the left column of 
  \figdot{low-e}.}
\end{figure}

This smoothening dynamics can be made more precise by plotting the logarithm
of the electron density along the symmetry axis of the streamer 
in Fig.~\ref{airXX_neg_cmp-sigma}. Photoionization creates a smoothly decaying
density tail ahead of the ionization front that initially is not visible
on a linear (non-logarithmic) scale. The point where the steep 
density decrease crosses over a smoother photoionization induced 
decay, moves toward higher density levels with time. For low fields
(\figdot{airXX_neg_cmp-sigma}, below)
up to the time when the streamer without photoionization branches, 
the large density levels visible in Fig.~\ref{low-e},
move essentially with the same velocity.
This is different in the high field case
(\figdot{airXX_neg_cmp-sigma}, above): 
there the photon created leading edge eventually dominates the
complete decay of the electron density and pulls the ionization front
to much higher velocities \cite{ebe2000}.

{\bf Acknowledgements:} A.L. was supported by the Dutch STW project 
CTF.6501, C.M. by the FOM/EW Computational Science program, both make 
part of the Netherlands Organization for Scientific Research NWO.


\end{document}